\documentclass[aps,prb,twocolumn,showpacs,floats]{revtex4}
\usepackage{graphicx}%
\usepackage{epsfig}
\usepackage{color}
\usepackage{amsmath}%
\usepackage{amssymb}%

\def\beq{\begin{equation}}

\def\eeq{\end{equation}}
\def\beqa{\begin{eqnarray}}
\def\eeqa{\end{eqnarray}}

\def\e{\epsilon}
\def\tb{\textcolor{black}}

\def\D{\Delta}

\def\ch{{\mathcal H}}

\def\hcon{\mathrm{H.c.}}
\def\r{{\bf r}}

\def\al{\alpha}

\def\si{\sigma}

\def\prl{{Phys. Rev. Lett.}\ }

\begin{document}

\bibliographystyle{prsty}


\title{\Large\bf Island formation in disordered superconducting thin films at finite magnetic fields}

\author{Yonatan Dubi$^1$\footnote{current address: department of physics, University of California-San Diego.}, Yigal Meir$^{1,2}$ and Yshai Avishai$^{1,2,3}$}
 \affiliation{
 $^{1}$ Physics Department, Ben-Gurion University, Beer Sheva 84105, Israel\\
 $^{2}$ The Ilse Katz Center for Meso- and Nano-scale Science and
 Technology, Ben-Gurion University, Beer Sheva 84105, Israel \\
 $^{3}$  RTRA researcher, CEA-SPHT (Saclay) and LPS (Orsay), France}.
\date{\today}

\begin{abstract} \noindent
\tb{It has been predicted theoretically and observed experimentally
that disorder leads to spatial fluctuations in the superconducting
(SC) gap. Areas where SC correlations are finite, coined SC islands,
were shown experimentally to persist into the insulating side of the
 superconductor-insulator transition.
 The existence of such (possibly weakly coupled) SC islands in  amorphous thin films of superconducting material } accounts for
  numerous experimental findings related  to superconductor-insulator transition and non-monotonic magneto-resistance
   behavior in the insulating region. In this work, a detailed analysis pertaining to the occurrence
of SC islands in disordered two-dimensional superconductors is presented.   Using a locally self-consistent
 numerical solution of the Bogoliubov-de-Gennes equations,
  the formation  of SC islands is demonstrated, and their evolution
with an applied perpendicular magnetic field is studied in some detail,
 along with the disorder-induced vortex-pinning.
 While mean-field theory cannot, in principle, explore phase
correlations between different islands,
 it is demonstrated that by inspecting the effect of a parallel magnetic field, one can show that
the islands are indeed uncorrelated SC domains. Experimental predictions based on this analysis are presented.

\end{abstract}

 \maketitle
\section{Introduction}
Interplay between disorder and superconductivity has been
 at the focus of attention for quite a long time. More than four decades ago,
  it has been established that the effect of weak disorder
on superconductivity  is not substantial.\cite{Anderson,Gorkov} Strong disorder,
however, may have a profound effect, driving the system from a superconducting (SC) to
an insulating state.  Such a superconductor-insulator transition (SIT) was
observed in two-dimensional amorphous SC films.
\cite{Goldman_review} Upon decreasing the film thickness or increasing a
perpendicular magnetic field, these films (which are held below
their bulk critical temperature) exhibit a transition from a SC
state, characterized by a vanishing resistance as $T \rightarrow
T_c$, to an insulating state, in which the resistance diverges as $T
\rightarrow 0$. \tb{The possibility} of tuning the system continuously between
these two extreme phases \tb{may be} a manifestation of a quantum phase
transition (which, strictly speaking, occurs at zero temperature).\cite{QPTreview}

Despite the substantial amount of experimental data, and numerous theoretical
investigations, there are still several unresolved issues pertaining to the physics of these
systems. One of the main puzzles is the mechanism by which the magnetic
field destroys the SC correlations. \tb{The "dirty boson" theory,\cite{Fisher,Girvin} describing bosons (Cooper pairs) in a random potential,}
 regards the SIT as a transition into a Bose-glass phase, in which the pairing amplitude is finite in the sample but
its phase is strongly fluctuating, giving rise to an insulating
state.
Such a phase is characterized by a pair-vortex duality, and its
hallmark is a universal value for the resistance $R_Q=h/4e^2$ at the
transition. While some experiments are consistent with that prediction,\cite{exps1,hebard,steiner} others are
not.\cite{exp2,Baturina} Beside the issue of universal resistance,
there are other observations which cannot  be simply explained by
 the dirty boson model, for instance, the temperature
dependence of the crossing point and the classical XY critical
exponent.\cite{Aubin}

The role of the large fluctuations in the local SC gap, $\D$, and
the formation of SC islands have been emphasized in several other
works. \cite{Spivak,Island:Galitzki,Island:Ghosal,Island:Semenikhin}
In Ref.~\onlinecite{Spivak} the authors predict the existence of
such islands by calculating the mesoscopic fluctuations of the
order-parameter at the mean-field level. Since phase fluctuations
are ignored by mean-field calculations, it was pointed out that
there is no real SC transition within that approximation. Later it has been shown
\cite{Island:Galitzki} that inclusion of quantum fluctuations
(beyond mean field) indeed results in a quantum phase-transition. A
detailed mean-field study of SC correlations in disordered
two-dimensional films has been presented in
Ref.~\onlinecite{Island:Ghosal}, employing a locally self-consistent
solution of the mean-field Bogoliubov-de Gennes (BdG) equations
\cite{Island:De-Gennes} at finite temperature and at zero magnetic
field. It was indeed found that disorder induces strong fluctuations
of $\D$ in space, resulting in regions with large order parameter
(here interpreted as SC islands - SCIs), separated by regions of
vanishingly
 small order parameter. The role of temperature is found \cite{Island:Semenikhin} to be similar to that
of disorder.


Recently we have  incorporated  thermal phase
 fluctuations beyond the BdG theory,\cite{us_nature} and have demonstrated
 the formation of isolated SCIs   within the two-dimensional disordered negative-U Hubbard model:
increasing magnetic field leads to the loss of
 SC phase correlations between different areas in the sample and an eventual SIT. The island picture is very helpful in explaining some of the experimental puzzles.
The coexistence and competition
 between Cooper pair and electron (or quasi-particle) transport give rise to the non-universality of the critical resistance.
It can also explain\cite{Island:us} the non-monotonic
magneto-resistance behavior observed on the insulating side of the
SIT.\cite{Shahar} The theory also predicts the persistence of the SC
islands into the insulating phase. Such SC correlations in the
insulating phase have been observed experimentally by using AC
transport (and measuring the superfluid stiffness on the insulating
phase) \cite{Crane} and by Aharonov-Bohm interference (where
oscillations with period $h/2e$ in the magnetoresistance were
observed in the magneto-resistance on the insulating phase,
indicating the presence of Cooper pairs). \cite{Valles}

In this picture,
the approach to the SIT from the insulating side is governed by the
emergence of phase correlations between the islands as temperature
or magnetic field are lowered, eventually resulting in a macroscopically
coherent sample. This picture is in agreement with the (quantum)
percolation scenario.\cite{Efrat,us_percolation} Evidence for percolation-like
behavior is substantiated in several
experiments.\cite{hebard,steiner,Baturina,Yazdani,DasGupta}
 Interestingly, the theory also demonstrated that phase correlations between different islands may
  be explored {\sl even within mean-field theory}, by looking at the response to parallel (Zeeman)
   magnetic field (see below). Thus, in principle, the SIT can also be investigated by solving the BdG equation in a finite magnetic field.
    In this work, however, we focus on the effects of magnetic field on the formation
and evolution of SC islands in highly disordered
two-dimensional SC systems in a wide range of  temperatures.
(We separate the effects of magnetic field into orbital field and Zeeman field, which we sometimes refer
 to as perpendicular and parallel fields, even though each of the latter leads to both orbital and Zeeman effects.)

The structure of this paper and the main achievements are listed
below. In section~\ref{model} the model and method of calculations
are briefly described. In section~\ref{Orbital} the role of
perpendicular magnetic fields is studied (assumed to act on the
orbital degrees of freedom as mentioned above). In particular, it is
found that, as the field is increased, the size of the SC islands
shrinks and the strength of the local order parameter diminishes. It
is explicitly shown that vortices are formed and their
disorder-induced pinning is analyzed. Moreover, these vortices tend
to accumulate in places where, in the absence of magnetic field, the
order parameter was low (that is in the "normal regions" of the
sample). The presence of these vortices suppresses the Josephson
couplings between the SCIs, leading eventually to completely
decoupled islands. Section~\ref{formation} is devoted to examining
the role of a parallel magnetic field, manifested
 by a Zeeman splitting. It is found that
at low temperatures and weak orbital magnetic fields, weakly disordered systems
exhibit an abrupt vanishing  of the SC order parameter  in response to increasing
parallel field, resembling a first order transition, as
expected for a uniform system.\cite{Island:CC} On the other hand,
when either temperature, disorder, or
 the orbital field are increased, the system undergoes a series of transitions, as a function of parallel field,  corresponding to the sequential vanishing of $\D$ on
 separate  islands, in agreement with  Ref.~\onlinecite{us_nature}.  The
emergence of local magnetic order in areas where superconductivity is
destroyed is demonstrated. In Section~\ref{summary} we summarize and discuss the results.

\section{Model and basic equations}\label{model}
 In this section, the model and method of calculations are briefly introduced.
The starting point is the effective tight-binding BdG Hamiltonian \cite{Island:De-Gennes}
on a square lattice,
 \begin{eqnarray} {\cal H}_{\rm BdG} =
-\sum_{<ij>,\sigma} t_{ij} ( c_{i\sigma}^{\dag} c_{j\sigma} + \hcon)
+\sum_{i} (\e_i-\tilde{\mu_i}) \hat{n}_{i\sigma} \nonumber\\
+ \sum_{i}[\Delta({\bf r}_i)c_{i\uparrow}^{\dag}c_{i\downarrow}^{\dag}
+ \Delta^{*}({\bf r}_i) c_{i\uparrow}c_{i\downarrow}]~~,
\label {eq:effhamil}
\end{eqnarray}
where the sum $<ij>$ runs on nearest neighbors,
$\e_i$ are random site energies taken from a uniform
distribution $[-W,W]$, with $W$ being the disorder strength,
$\tilde{\mu_i}=\mu+\frac{|U|}{2}\langle n_i \rangle$ is the local
 Hartree-shifted chemical potential,  \beq \D ( \r_i )=-|U| \langle
c_{i\downarrow}c_{i\uparrow}\rangle \eeq is the local pairing
potential and $U<0$ is an effective electron-electron on-site
attractive interaction.\cite{Island:De-Gennes} The orbital magnetic
field enters via the nearest-neighbor hopping integral $t_{ij}$. In
the Landau gauge it can be written within the Peierls substitution
rule as  \beq t_{ij}=t_0 \exp (-i \al
y_{i}) \label{hopping} \eeq if $\r_i-\r_j$ is in the ${\bf
x}$-direction, and $t_{ij}=t_0$
 otherwise.  In Eq.~\eqref{hopping} $\al=\phi / \phi_0$ where $\phi$ is the magnetic flux per plaquette,
$\phi_{0}=h c/e$ is the flux quantum, and $y_i$ is the ${\bf
y}$-direction coordinate (in units of lattice spacing). Hereafter,
$t_0$ is set to unity and serves as reference to all other energy
scales. Using a Bogoliubov transformation one can diagonalize
$\ch_{\rm BdG}$ exactly and obtain the BdG equations, \beq \left(
\begin{array}{cc}
 \hat{\xi} & \hat{\D} \\
 \hat{\D}^{*} & -\hat{\xi}^{*} \\
\end{array}%
\right)
\left(
\begin{array}{c}
  u_n(\r_i)  \\
  v_n(\r_i)  \\
\end{array}%
\right)=E_n
\left(
\begin{array}{c}
  u_n(\r_i)  \\
  v_n(\r_i)  \\
\end{array}%
\right) ~~. \label{eq:bdg} \eeq In Eq.~\eqref{eq:bdg},
$\hat\xi u_{n}({\bf r}_i) = -\sum_{\hat\delta} t_{i,i+\delta}
u_{n}({\bf r}_i+\hat\delta)+(\e_i-\tilde{\mu_i}) u_{n}({\bf r}_i)$
where $\hat\delta = \pm{\hat{\bf x}}, \pm{\hat{\bf y}}$, and
$\hat\Delta u_{n}({\bf r}_i) = \Delta({\bf r}_i) u_{n}({\bf r}_i)$
and similarly for $v_{n}({\bf r}_i)$, and the energies are the quasi-particle (QP)
excitation energies $E_n \geq 0$. The pairing potential $\D(\r_i)$ and
the electron density per site $n_i$ are given, in terms of the QP amplitudes $u(\r_i)$ and $v(\r_i)$,
\begin{eqnarray} \Delta({\bf r}_i) &=& |U(\r_i) |\sum_{n}u_{n}({\bf
r}_{i})v_{n}^{*}({\bf r}_{i})
\nonumber\\
\langle n_i \rangle &=& 2 \sum_{n}|v_{n}({\bf r}_{i})|^{2} ~~.
\label {eq:selfc} \end{eqnarray}
\par In what follows, Eq.~(\ref{eq:bdg}) is solved numerically on a
finite (albeit large) two-dimensional square lattice with
 hard-wall boundary conditions,
 starting with
some initial guesses for $\Delta({\bf r}_i)$ and $n_i$.
 $\Delta({\bf r}_i)$ and $n_i$ are then determined via Eq.~\eqref{eq:selfc},
  and the procedure is repeated until self-consistency is obtained {\sl on each site}. In order to be able to compare the same system at
 different magnetic fields,
the chemical potential is altered in every iteration such that the total average
 density per site $\langle n \rangle =\frac{1}{N} \sum_i n_i$ remains constant
($N$ being the total number of sites in the lattice).

\section{Numerical results}\label{numres}
\subsection{Effect of orbital magnetic field}\label{Orbital}
We start by examining the effect of an orbital magnetic field on the
pairing potential $\D$. In Ref.~\onlinecite{Island:us} it was
argued that the SCIs diminish in size and may even vanish
as the magnetic field is increased. This assumption is indeed
confirmed by the numerical results.
Figure~\ref{D(R,B)} depicts the spatial distribution of the order
parameter amplitude, $|\D(\r)|$, in the two-dimensional system, for
three different values of the magnetic field.
  At
zero magnetic field the order parameter fluctuates in space due to
disorder, assuming larger values (bright areas in
 Fig.~\ref{D(R,B)}) in places where the effective local potential is small, i.e., $\e_i-\tilde{\mu_i} \approx 0$,
\cite{Island:Ghosal,Island:Herbut} which facilitates fluctuations
between 0 and 2 electrons on the site.
 These areas are separated by regions with small order parameter (dark areas in
Fig.~\ref{D(R,B)}) and hence constitute "superconducting islands"
(as will be shown in following sections). As the magnetic field is
increased, the size of these islands and the strength of the order
parameter diminishes,  and some SCIs virtually  disappear (that is,
at very strong fields the order parameter there becomes smaller than
the numerical tolerance). It should be pointed out that while the
average value of the order parameter diminishes by nearly $50 \%$
when the flux is increased from $\phi/\phi_0=0$ to
$\phi/\phi_0=0.2$, the maximal value of the order parameter drops by
less than $10\%$, meaning that the concept of "SC islands" persist
well within the strong field region (it is the loss of correlations
between the islands that drives the system into its normal state).
\begin{center}
\begin{figure}[h!]
\centering
\includegraphics[width=9truecm]{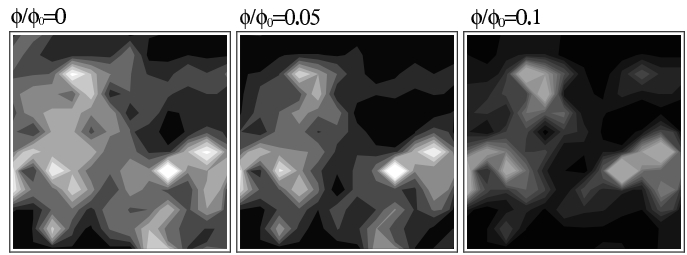}
\centering \caption{\footnotesize Spatial distribution of the order
parameter amplitude $|\D|$ for different values of the orbital
magnetic field, $\phi/\phi_0=0,0.05,0.1$. Bright regions,
corresponding to large values of $|\D|$, constitute "superconducting
islands", separated by regions of small $|\D|$ (dark regions).
Increasing magnetic field leads to attenuation of the SC order
parameter and even to the vanishing of SC order in some areas of the
sample. Calculation is performed on a $12 \times 12$ size lattice
with electron density $\langle n \rangle=0.875$, disorder strength
$W=1$ and interaction strength $U=2$ (this value for $U$ is
maintained throughout).} \label{D(R,B)} \end{figure}
\end{center}

\begin{figure}[h!]
\centering \vskip 1truecm
\includegraphics[width=8truecm]{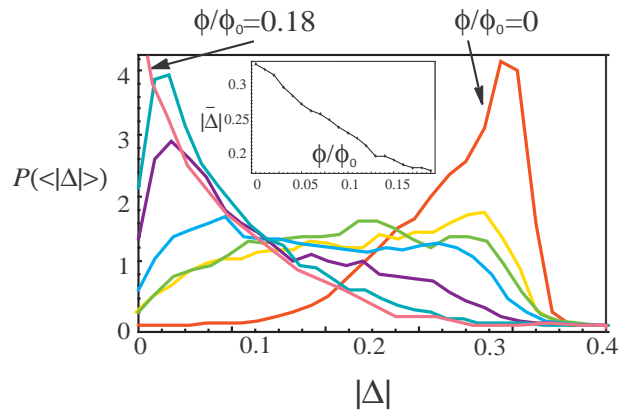}
\centering \caption{\footnotesize (Color online) Distribution
function of the SC order parameter amplitude $|\D|$, for seven different
values of magnetic field, $\phi/\phi_0=0-0.18$ (calculated for a
system size $16 \times 16$, averaged over the entire lattice and
over $25$ realizations of disorder). With increasing
magnetic field the distribution
 shifts from being peaked at finite value
 (the uniform-system value) to being peaked
 at zero value. Note, however, the persistence of tails at large $|\D|$
 even at high magnetic field, indicating the existence of regions with large $|\D|$. Inset: the average order parameter
amplitude, $\langle |\D|\rangle$, as a function of $\phi/\phi_0$,
showing a decrease in the SC order parameter with the applied orbital
magnetic field.
}
\label{D(B)} \end{figure}

The distribution function of the order parameter amplitude,
$P(|\D|)$, is plotted in Fig.~\ref{D(B)} for  different values of
the magnetic field. As can be seen, the distribution function shifts
from being peaked around some finite initial value to being peaked
at $\D=0$. However, the distribution tails persist even at
high-field, indicating that SC islands are present.

In the inset of Fig.~\ref{D(B)} the average order parameter
amplitude $\langle |\D|\rangle$ is plotted as a function of
$\phi/\phi_0$. We note that (i) in a finite lattice $\langle
|\D|\rangle$ never strictly vanishes and (ii) for a given disorder
realization, $\langle |\D|\rangle$ may be a non-monotonic function
of $\phi/\phi_0$. In fact, in a clean system $\langle |\D|\rangle$
has been shown to
 oscillate as a  function
of magnetic field.\cite{Island:Maska} Disorder, however, tends to smear the non-monotonicity, and
upon averaging over disorder realizations one
 obtains a monotonically decreasing function.

\begin{center}
\begin{figure}[h!]
\centering
\includegraphics[width=9truecm]{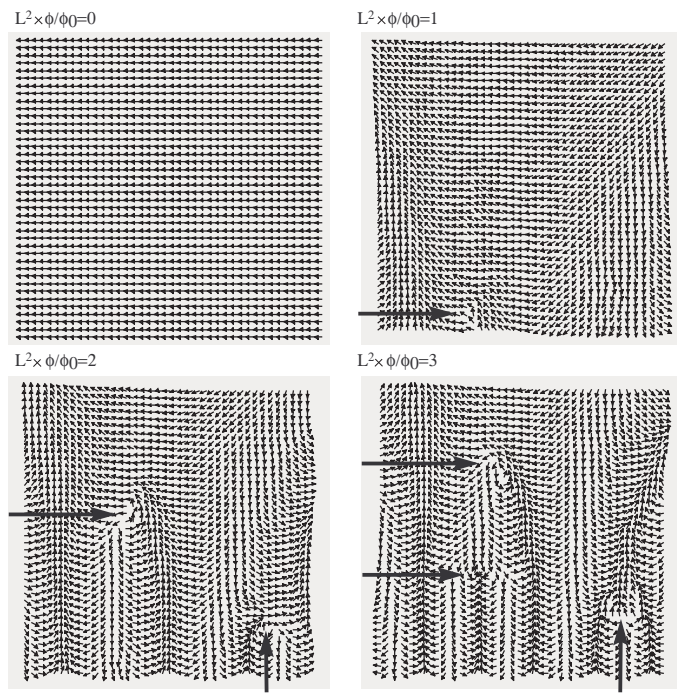}
\centering \caption{\footnotesize Spatial mapping of the SC order
parameter phase for values of magnetic field corresponding to a
different number of flux quanta in the system  (i.e.,
$\phi/\phi_0 \times (L \times L) =0,1,2,3$), on a $40 \times 40$
size sample. Arrows point at the
positions of the vortices. For each number of flux quanta, a similar
number of vortices is formed, their position is irregular and
determined by the disorder potential.
 }
\label{vortices0} \end{figure}
\end{center}

Another illustration of the interplay between  disorder and magnetic
field is manifested in the location of the vortices. In a clean system
one expects the vortices to form a regular Abrikosov lattice. However,
strong disorder modifies the position of vortices, as they are pinned
within regions where the amplitude of the order parameter has a small value.
In Figure~\ref{vortices0} the phase of the SC order parameter is
plotted for different values of the magnetic field corresponding to the consecutive
insertion of one additional flux quantum into the system.  As can be
seen, a new vortex enters the system each
 time an additional flux quanta is inserted, and vortices do not form a lattice. Rather,
  their position is affected by disorder.

\begin{figure}[h!]
\begin{center}
\includegraphics[width=9truecm]{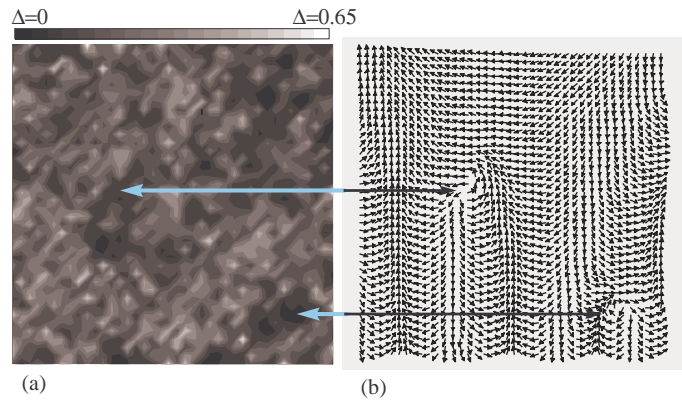}
\centering \caption{\footnotesize Comparison between (a) the order
parameter amplitude at zero field and (b) the phase of the order parameter
at finite field (two flux quanta), indicating that the vortices are
formed at the position of weak superconducting order  (dark regions in (a)). }
\label{vortices}
\end{center}
\end{figure}

These correlations can be exposed more clearly. In
Figure~\ref{vortices} the order parameter amplitude at zero field is
plotted (Fig.~\ref{vortices}(a)), together with the phase of the order parameter
 (Fig.~\ref{vortices}(b)), for the same system as in
Figure~\ref{vortices0} with two flux quanta. As can be seen, the vortex
cores are positioned where the amplitude is low (dark regions in
(a)), i.e., they are "pinned" to
 regions with low order parameter. The emerging interactions between vortices still play a significant
  role, as, for example, the second vortex that enters the system changes the position of the first one.

  In Ref.~\onlinecite{us_nature} it was shown that the SIT is
driven by loss of phase-coherence between the SC phases on different
islands. The fact that vortices appear in regions of low order
parameter amplitude, i.e. in between the SC islands, is consistent
with the picture that orbital magnetic field causes loss of phase
coherence between the islands: the effective Josephson coupling
between the islands is reduced with increasing magnetic field, until
the Josephson energy becomes smaller than temperature
(or the amount of quantum phase
fluctuations) and the islands decouple.

\subsection{Evolution of superconducting islands  in a parallel magnetic field}\label{formation}
In the previous subsection the roles of disorder and orbital
magnetic field in generating fluctuations in the amplitude of the SC
order parameter have been demonstrated. However, the question remains
whether these fluctuations may be regarded as separated SC islands.
Here we demonstrate how the Zeeman effect resulting from an application of a parallel
magnetic field helps to answer this
question. This is carried out by
extending our previous calculations,\cite{us_nature} and include the electron spin
through the Zeeman effect.

When a parallel magnetic field is applied, the single electron levels split.
It was shown a long time ago \cite{Island:CC}
 that upon applying  a parallel field, a uniform superconductor undergoes a first-order transition.
 The reason for this transition is simple: superconductivity requires occupation of paired spin up and spin down states, which becomes costly with increasing Zeeman energy.
  For a single pair, one expects the transition to take place when the gain
 from superconductivity, $\D_0$, (the bulk value of the
order parameter), is overcome by Zeeman energy, $h_c\equiv g\mu_B B_{||}$, where $B_{||}$ is the strength of the parallel field,
  $g$ being the g-factor for electrons and
$\mu_B$ is the Bohr magneton. In fact, when correlation effects are taken into account, one finds that the
  transition from superconductivity to a magnetically aligned state occurs at $h_c=\D_0 /\sqrt{2}$.  For ultra-small isolated SC grains it was shown \cite{Island:Braun} that a finite size effect (manifested in a level spacing of the order of $\D_0$) may smooth the otherwise sharp transition.

In order to account for the effect of parallel field, a Zeeman splitting term
 $\sum_{i \si} \si h c^\dagger_{i \si}c_{i \si}$ is added to the Hamiltonian of Eq.~\eqref{eq:effhamil}. The
Bogoliubov transformation now has to be performed with spin-dependent quasi-particle amplitudes
 $u_{n \pm}(\r_i)$ and  $v_{n \pm}(\r_i)$, which now obey the spin-generalized BdG equations,
\cite{Island:Book2}
\beq
\left(
\begin{array}{cc}
 \hat{\xi} \pm h  & \hat{\D} \\
 \hat{\D}^{*} & -\hat{\xi}^{*} \pm h \\
\end{array}%
\right)
\left(
\begin{array}{c}
  u_{n \pm}(\r_i)  \\
  v_{n \pm}(\r_i)  \\
\end{array}%
\right)=(E_n \pm h) \left(
\begin{array}{c}
  u_{n \pm}(\r_i)  \\
  v_{n \pm}(\r_i)  \\
\end{array}%
\right) ~~. \label{eq:bdg_spin} \eeq Once these equations are
solved, the local pairing potential and occupation are determined by
 self-consistent equations similar to that of Eq.~\eqref{eq:selfc}.
Note that for singlet (S-wave) pairing, the different
spin-generalized quasi-particle amplitudes are decoupled, and the
coupling between the different spins enters only in the
determination of the self-consistent Hartree shift and order
parameter.

For a clean or weakly disordered system and for parallel field strength $h<h_c$, the
eigenvectors are independent of $h$ (due to the
simple dependence of the energies on the Zeeman term), and hence the solution for the
pairing amplitude $\D(\r)$ and the local occupation is unchanged. On
the other hand, for $h>h_c$ a solution with $\D(\r)=0$ yields a
lower free energy. Thus, one should observe a step-like behavior of
the order parameter. In contrast, in a disordered system, as
$\D(\r)$ varies in space, one also expects $h_c$ to be
space-dependent. For a system composed of separated SCIs, one
thus expects that the order parameter on each island will vanish
when $h>h_c(i)$ (where $i$ is an index for identifying the corresponding island). Thus, the
vanishing of the average value of $\D$ is expected to exhibit multiple steps, each step
corresponds to the vanishing of superconductivity in a different
island.

\begin{figure}[h!]
\centering
\includegraphics[width=8truecm]{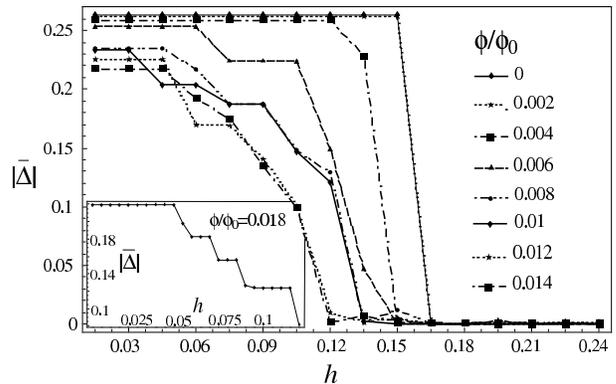}
\centering \caption{\footnotesize The average order parameter
amplitude $\langle |\D|\rangle$ as a function of the Zeeman
splitting $h$ for different values of orbital magnetic field,
$\phi/\phi_0=0-0.014$. For small $\phi/\phi_0$, the system undergoes
a first-order transition, which is smoothed as $\phi/\phi_0$
increases. For $\phi/\phi_0=0.004$ steps in $\langle |\D|\rangle$
are visible. Inset: the same quantity for $\phi/\phi_0=0.018$, but with
larger resolution in $h$. At this resolution, steps in the $\langle
|\D|\rangle$ are clearly seen.  }
\label{D_parallel} \end{figure}

In Figure~\ref{D_parallel} the average order parameter amplitude
$\langle |\D|\rangle$ is plotted as a function of $h$ for different
orbital
 fields. One can clearly see that for small $\phi$ (weak
 perpendicular magnetic field), the system undergoes
 a single sharp
transition.  In contrast, for finite values of $\phi$, step
structure of $\langle |\D|\rangle$ is indeed visible. This is seen
more clearly in the inset of Figure~\ref{D_parallel} where $\langle
|\D|\rangle$ is plotted as a function of $h$ at a perpendicular
field $\phi/\phi_0=0.018$, with much higher resolution in $h$. At
this resolution, the step-like behavior is seen even for this high
value of the orbital field. Thus, large orbital field suppresses SC
correlations between the islands. They seem to lose SC order
separately and successively,  depending on the specific value of the
gap for each SCI.

\begin{center}
\begin{figure}[h!]
\centering
\includegraphics[width=9truecm]{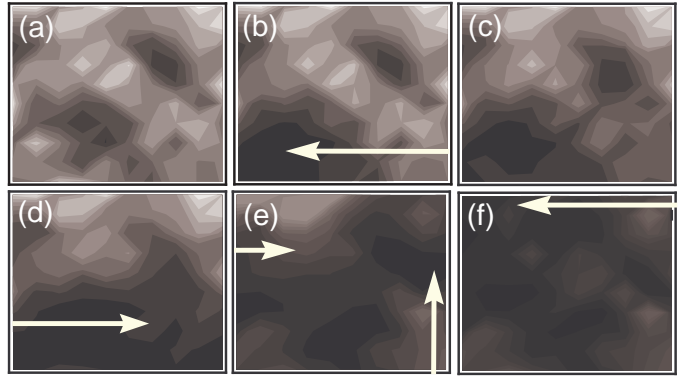}
\centering \caption{\footnotesize Spatial distribution of the order
parameter amplitude $|\D|$ at $\phi/\phi_0=0.014$ for $h=0-0.12$
(a-f). Arrows indicate regions in which the order parameter vanishes
abruptly (SC islands).}
\label{D(R)_BZ2} \end{figure}
\end{center}

To make this visual, the spatial distribution of the pairing
parameter at $\phi/\phi_0=0.014$ is plotted  in
Figure~\ref{D(R)_BZ2} for $h=0-0.012$ (Fig.~\ref{D(R)_BZ2}(a-h)). The consecutive vanishing of the order parameter in different areas with increasing parallel magnetic field is clearly visible. It
is noticeable that even when the order parameter vanishes, as a
result of the field, in one area of the sample, the
other regions remain almost unaffected. The different regions (for
which the order parameter vanishes at different parallel fields) can
thus be indeed regarded as separated SCIs.

\begin{center}
  \begin{figure}
  \centering
  \includegraphics[width=9truecm]{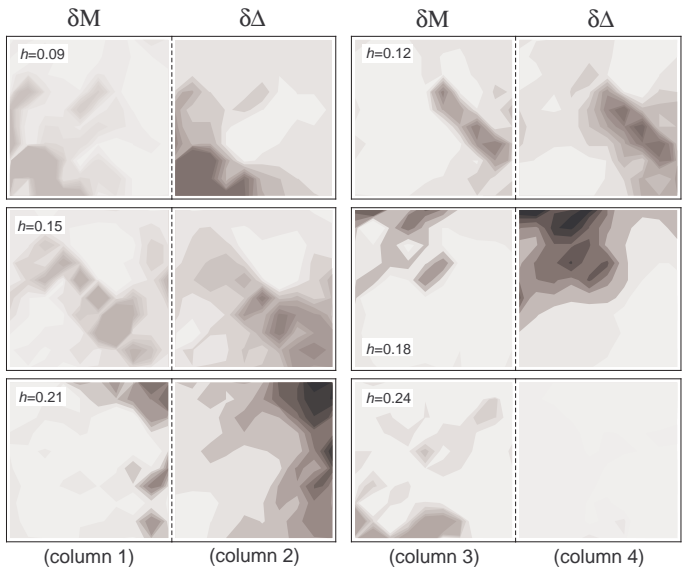}
  \centering
  \caption{\footnotesize Spatial distribution of
  the differential changes in magnetization, $\delta M(\r,h)\equiv M(\r,h)-M(\r,h-0.03)$ (columns 1 and 3), and in pairing potential amplitude, $\delta (\r,h)\equiv\D(\r,h)-\D(\r,h-0.03)$ (columns 2 and 4), for different
   Zeeman fields $h=0.045-0.12 $, and for orbital field $\phi/\phi_0=0.014$. The emergence of
   magnetic correlation occurs simultaneously and is highly correlated with the vanishing of SC order.}
  \label{Magnetization_BW}
 \end{figure}
  \end{center}

As already mentioned, another indication
for the destruction of the SC order may be elucidated by examining
the local magnetization
 $M(\r)=\langle n_{\uparrow }(\r)-n_{\downarrow }(\r) \rangle$,
 which develops locally where SCIs are
  destroyed by the parallel field.
  This is shown in Figure~\ref{Magnetization_BW}, where we plot side by side the spatial distributions
of the
 {\sl differential variations} of the magnetization and of the pairing potential, for a finite orbital field and for different Zeeman fields (same values as in Fig.~\ref{D(R)_BZ2}).
 It is clearly seen that the magnetization
 (dark regions in the left panels) develops only where SCIs  (dark regions in the right panels) shrink. Notice that even when the parallel field strength exceeds the critical value ($h_c \approx 0.24 $ for these parameters), the magnetization still
fluctuates, now due to correlations with the local electron density.

An important and experimentally testable prediction of the above calculation is that a strongly
disordered SC film subject to a parallel field may have both a zero
resistance state and finite magnetization. This is in contrast with
a clean film, which is either SC or spin polarized.

\section{Summary and discussion} \label{summary}
In this paper the physics related to formation of SCIs in disordered
two-dimensional SC films and their evolution under an applied magnetic field
was addressed. SCIs are defined as specific domains exhibiting  large value of the order parameter
amplitude surrounded by other domains for which the amplitude of the order parameter is small.
Using the locally
self-consistent solution of the BdG equations, it was demonstrated
that SCIs are formed in disordered SC films. Upon an
increase of the orbital magnetic field, the size of these islands tend to decrease,
together  with the strength of the order parameter. This is accompanied  by  loss of SC phase correlations between different islands.

In order to verify that SCIs indeed constitute separate isolated domains, the application of
parallel magnetic fields has been included in our  numerical
simulations. For a uniform or weakly disordered sample it results in a
 first order transition into a normal state.
 In contrast, for medium and high disordered systems and in the presence of an orbital magnetic field,
 it is found that upon increasing the
 parallel magnetic field,
the average order parameter vanishes in a series of steps,
indicating that each island undergoes a phase transition at its
own turn, with its own critical parallel field. This substantiates the picture
of well separated and isolated SCIs. In Ref.~\onlinecite{us_nature} we have demonstrated a one-to-one correspondence between the phase correlations between the islands, at finite orbital field and zero Zeeman field, on one hand, and their sequential response to the Zeeman magnetic field, on the other hand. This suggests that the SIT can be studied within mean-field theory, even though the latter does not include phase fluctuations.


The above results can in principal be addressed experimentally,
 by means of local measurements using, e.g., scanning tunneling microscopy, which is also sensitive to magnetic
 order. We predict that, as a function of parallel magnetic field,  the (average) SC order parameter will exhibit a behavior resembling that of Figure~\ref{D_parallel}, that is the transition
will be smoothed when the orbital effects are dominant. The local SC gap on
an isolated island, on the other hand, should exhibit a sharp
attenuation characteristic of a first-order transition. This will be accompanied by local formation of magnetic order, so we expect local magnetism coexisting with either bulk superconductivity (if the SCIs percolate) or local superconductivity (if they do not).

Lastly we point out that some of the phenomena discussed here were also observed in high-$T_c$ superconductors. Local variations in the SC gap have been observed by STM measurements.\cite{STM-highTc} In the underdoped region, it was observed \cite{Gomes} that local SC gap persists well into the normal phase. Even more relevant is the observation \cite{Sonier} in these materials of inhomogeneous magnetic response, which persists above $T_c$. Thus disorder may play an important role in high-$T_c$ superconductors, especially in the underdoped regime, leading, for example, to the interpretation of the pseudo-gap, as an average over the spectra of SC and non-SC regions.

\begin{acknowledgments}
This research was partially supported by a grant from the Israeli
Science Foundation (ISF). YD acknowledges financial support from the
Kreitmann Foundation Fellowship. YM thanks the Aspen Center of Physics. We thank A. Sharoni for fruitful
discussions.
\end{acknowledgments}

\end{document}